\documentclass[aps,twocolumn,floatfix,showpacs]{revtex4}

\usepackage{graphicx}
\usepackage{scrtime}
\usepackage{latexsym}

\usepackage{epsfig}
\usepackage{fancyhdr}
\usepackage[hang]{subfigure}
\usepackage{supertabular}
\usepackage{wrapfig}
\usepackage{amssymb}

\begin{document}

\title{Hopping conduction in strongly insulating states of a diffusive bent quantum Hall junction}

\author{L. Steinke$^1$, D. Schuh$^{1,2}$, M. Bichler$^1$, G. Abstreiter$^1$, and M. Grayson$^{1,3}$}
\affiliation{
$^1$Walter Schottky Institut, Technische Universit\"at M\"unchen, D-85748 Garching, Germany\\
$^2$Universit\"at Regensburg, Institut f\"ur Angewandte und Experimentelle Physik II, Universit\"atsstrasse 31, D-93040 Regensburg, Germany\\
$^3$Department of Electrical Engineering and Computer Science, Northwestern University, Evanston, IL 60208, USA}

\begin{abstract}
Transport studies of a bent quantum Hall junction at integer filling factor $\nu$ show strongly insulating states ($\nu$ = 1,2) at higher fields. In this paper we analyze the strongly insulating behavior as a function of temperature $T$ and dc bias $V_{dc}$, in order to classify the localization mechanisms responsible for the insulating state.
The temperature dependence suggests a crossover from activated nearest-neighbor hopping at higher $T$ to variable-range hopping conduction at lower $T$ ($G \sim \exp(-(\frac{T_{0}}{T})^{1/2})$). The base temperature electric field dependence shows $I({\cal E})\sim \exp(-(\frac{{\cal E}_0}{\cal E})^{1/2})$, consistent with 1D variable-range hopping conduction. We observe almost identical behavior at $\nu = 1$ and $\nu = 2$, and discuss how the bent quantum Hall junction conductance appears to be independent of the bulk spin state.  Various models of 1D variable range hopping which either include or ignore interactions are compared, all of which are consistent with the basic model of disorder coupled counter-propagating quantum Hall edges.  
\end{abstract}
\maketitle
Disorder in one-dimensional (1D) conductors can backscatter propagating charge, leading to localization of electronic states at low temperatures.\cite{GangOfFour}
Interactions could further modify the conduction characteristics \cite{SHK,Fogler_hopping,Nattermann} and lead to a temperature-dependent localization transition \cite{GOR, Basko}.
In chiral 1D systems like quantum Hall (QH) edges \cite{Halperin,Wen} charge propagates in only one direction, but with counter-propagating chiral quantum Hall edge states coupled along an extended junction \cite{KANG, 1, Matt}, a nonchiral interacting 1D system can form at such a QH line junction \cite{RENN,KANE,KANE_FISHER,kim,kollar,mitra}, and reveal information about generic 1D systems.

In this work we study an extended  line junction of integer quantum Hall edges in a 90$^{\circ}$ bent QH (BQH) junction. 
Previous work introduced the corner overgrowth technique for fabricating bent two-dimensional electron systems (2DES) and demonstrated the propagation of quantum Hall edge channels along the 90$^{\circ}$ junction \cite{1}.
At equal filling factors $\nu$ on both facets of the bent quantum Hall system, an equal number of edge channels propagates along the junction in both directions, forming a non-chiral system of strongly coupled edge states.
Conductance measurements of the bent quantum Hall junction show evidence for two different phases in the integer quantum Hall regime, depending on the filling factor $\nu$: A strongly insulating phase at $\nu=1$ and $\nu = 2$ and a weakly insulating phase at integer $\nu \ge 3$ \cite{Matt}.  This paper concerns itself with classifying the localization mechanisms that lead to an insulating state at the lowest Landau level filling factors.

The investigated sample structure is a high-mobility two-dimensional electron system bent at a 90$^\circ$ angle, fabricated using corner overgrowth \cite{1,Matt,footnote}. A modulation doped AlGaAs/GaAs heterointerface structure is overgrown  
\begin{figure}[!ht]
\center
\includegraphics[width=8cm]{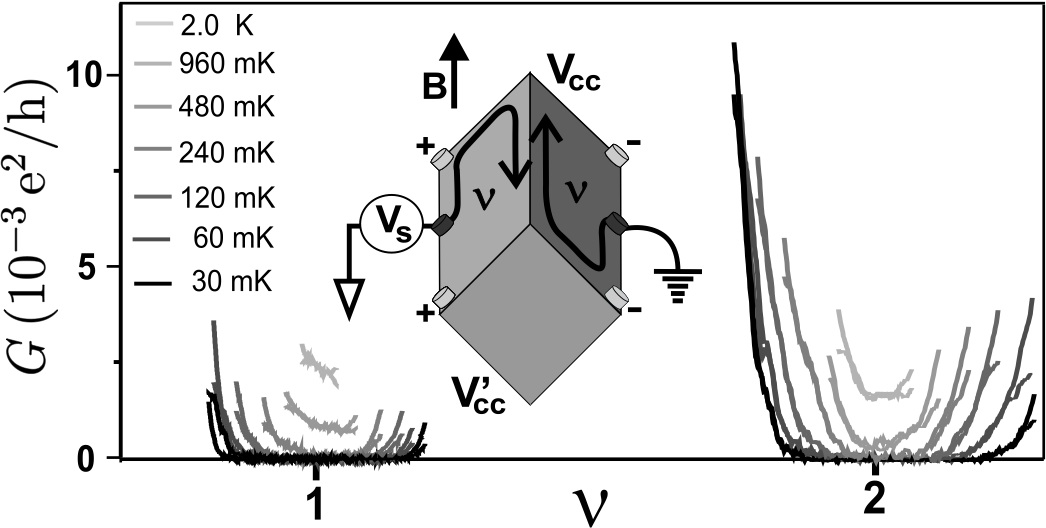}
\caption{Small signal conductance $G=dI/dV \vert_{V=0}$ of the bent quantum Hall junction vs. filling factor $\nu$ and temperature. The inset illustrates the flow of edge currents in the bent quantum Hall system, for the case of equal $\nu$ on the two facets. The current flowing along the bent quantum Hall junction is proportional to the cross-corner voltage drops $V_{cc}$ or $V'_{cc}$, respectively. With the voltage $V_{s}$ between the edge channels incident to the junction, the conductance $G$ along the junction can be calculated. The gray scale indicates temperature increments by $\times 2$.} 
\label{Gvsnu}
\end{figure}
on an {\it ex-situ} cleaved GaAs (110) corner substrate, where a bent 2DES forms at the interface between the
GaAs base layer and the AlGaAs barrier. The exact layer sequence is described in Ref. \cite{1}. Indium contacts alloyed along the edges of both facets of the sample provide ohmic contacts to the bent 2D system.
After illumination with red light at $T = 4.2$ K we obtain electron sheet densities of $n_{(110)} = 1.08\cdot 10^{11}\,{\rm cm}^{-2}$ on the (110)-facet and $n_{(1\bar{1}0)} = 1.37\cdot 10^{11}\,{\rm cm}^{-2}$ on the $(1\bar{1}0)$-facet, with minor variations for each illumination and cooldown. The estimated electron mobility on both facets is $\mu = 0.5 \cdot 10^6 \,{\rm cm^2/Vs}$. The length of the junction is approximately 4 mm, and length dependence measurements in a previous publication \cite{Matt} have shown that the conduction along an extended junction is diffusive with multiple scattering events over such macroscopic lengths.
\begin{figure}[!ht]
\center
\includegraphics{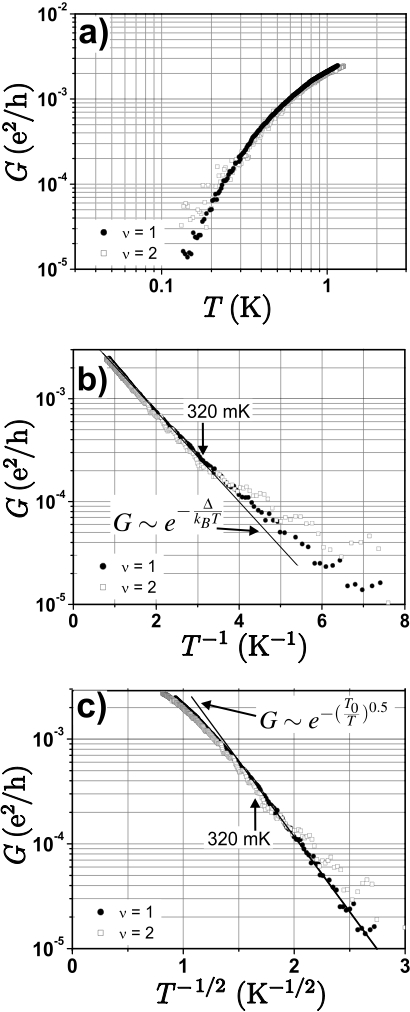}
\caption{a) Small signal conductance $G=dI/dV\vert_{V=0}$ at filling factors $\nu = 1,2$, plotted on a double-log scale vs. temperature $T$. We observe a very low conductance with strong temperature dependence. Panel b) shows an Arrhenius plot of the same conductance data $G$ at $\nu = 1$ and $\nu = 2$. In panel c) the same data is plotted semi-logarithmically vs. $T^{\,-\frac{1}{2}}$. The low temperature behavior (below $\approx 320$ mK) fits an exponential in $T^{\,-\frac{1}{2}}$, consistent with one-dimensional VRH conduction. At higher temperature the conductance seems exponential in $(1/T)$, indicating a crossover to activated behavior.}
\label{Tdep}
\end{figure}
We measure the conductance of the bent quantum Hall junction, with a tilted magnetic field $B$ applied at a tilt angle such that both facets form quantum Hall systems with equal filling factors, Fig. \ref{Gvsnu} inset. 

As developed in Ref. \cite{Matt}, the conductance $G$ is\cite{RENN,KANE}:
\begin{equation}
G=\nu \frac{e^2}{h}\frac{V_{cc}}{V_s}=\nu \frac{e^2}{h}\frac{V_{cc}'}{V_s}
\label{Gwire}
\end{equation}
provided both facets are in the QH regime, \(R_{xx} = 0\), such that no current is scattered away from the junction ($V_{cc} = V'_{cc}$). Fig. \ref{Gvsnu} shows a plot of the BQH junction conductance measured at various filling factors $\nu$ and at fixed temperatures between 30 mK and 2.0 K. The two curves shown for each temperature correspond to the conductance deduced from a measurement of $V_{cc}$ and $V_{cc}'$ and are plotted over the full range where they overlap each other.  Note how at higher temperatures the filling factor interval narrows for measuring junction conductance.

The conductance clearly vanishes at the base temperature $T = 30$ mK, and rises strongly with increasing $T$. Phenomenologically we label the behavior at $\nu = 1,2$ as strongly insulating. To measure the temperature dependence of the conductance, the magnetic field is set at $\nu = 1$ or $\nu = 2$, respectively, while the conductance along the BQH junction is measured in a continuous temperature sweep. Fig. \ref{Tdep} a) shows a double-logarithmic plot of the zero bias conductance $G = dI/dV\vert_{V = 0}$ vs. temperature $T$ measured with a 10 nA AC excitation. As the temperature is decreased by roughly one order of magnitude, the conductance $G$ in Fig. \ref{Tdep} a) drops by two orders of magnitude, where no single exponent is able to characterize the full temperature range. Note the increased noise level in the $\nu = 2$ data compared to $\nu = 1$, which is due in part to the factor of $\nu = 2$ lower signal to noise inherent in Eq. \ref{Gwire}.

To investigate whether there is evidence of activated nearest-neighbor hopping conduction, the data from Fig.~\ref{Tdep}(a) is plotted in an Arrhenius plot in Fig.~\ref{Tdep}(b).  The high temperature data above $\sim 320$ mK is
fitted to an activated temperature dependence $G \sim \exp(\frac{-\Delta}{k_BT})$, with $\Delta = 87.3 \, \mu$eV at $\nu = 1$ and $\Delta = 89.0 \, \mu$eV at $\nu = 2$.  To check for variable range hopping conduction, the same data is plotted in Fig.~\ref{Tdep}(c)  vs. $T^{-1/2}$. At temperatures below $\sim 320$ mK the conductance can be
described by variable-range hopping conduction of the form $G(T) \sim \exp(-(\frac{T_0}{T})^{1/2})$ with $T_0 = 11.0  $ K
at $\nu =1$. The data at $\nu = 2$ is consistent with $\nu = 1$, yet the higher noise level does not allow for an independent fit. These observations indicate crossover from thermally activated nearest-neighbor hopping conduction at higher temperatures to
one-dimensional VRH conduction at lower $T$ \cite{ALE04, Shante, Kurki}.

\begin{figure}[!ht]
\center
\includegraphics{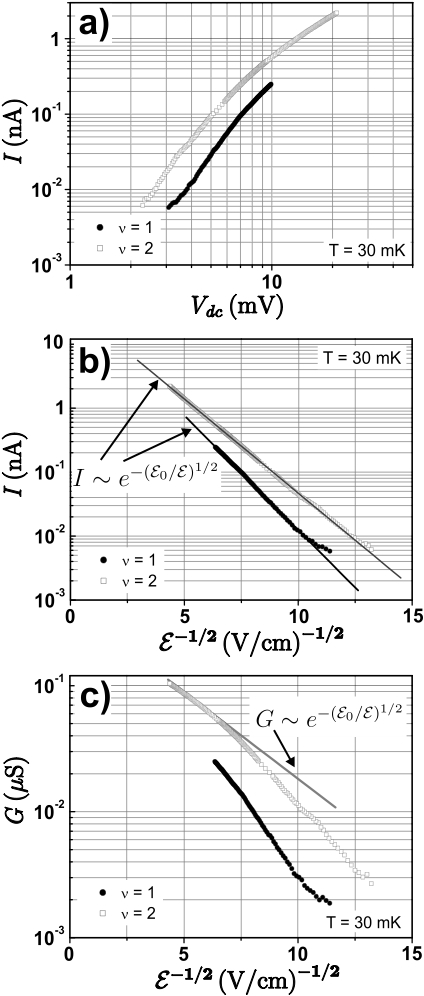}
\caption{(a) The $I(V)$ curves at base temperature $T = 30$ mK, plotted on a double-log scale.
(b) The current $I$ plotted semi-logarithmically vs. the inverse square root of the applied dc electric field ${\cal E}$.  For $\nu = 1$ ($\nu = 2$), an exponential in ${\cal E}^{-1/2}$ describes the $I({\cal E})$ dependence quite well. (c) Large signal conductance $G = I/V$ also versus inverse square root of  ${\cal E}$.  An exponential dependence $G({\cal E}) \sim \exp(-({\cal E}_0/{\cal E})^{1/2}$ fits only the $\nu = 2$ data, and only at the highest electric fields.}
\label{Vdep}
\end{figure}

Since the characteristics of VRH conduction should also become evident in the electric field dependence of the conductance, we measure the differential conductance along the BQH junction as a function of an applied dc bias. For these measurements we apply a dc bias voltage $V_{dc}$ modulated with an ac signal of $V_{ac} = 20 \, \mu$V to the $V_s = V_{dc} + V_{ac}$ contacts sketched in Fig. \ref{Gvsnu}. The ac current along the junction was again measured by the cross corner voltages $V_{cc}$ and $V'_{cc}$ with a known $R_{xy}$ facet impedance.
Fig. \ref{Vdep} a) shows a double-logarithmic plot of the $I(V_{dc})$ curves at $\nu = 1,2$ obtained by integrating the differential conductance over $V_{dc}$. Compared to the temperature dependence shown in Fig. \ref{Tdep}, the $I(V_{dc})$ curves of Fig.~\ref{Vdep} show a similarly strong dependence on $V_{dc}$, where $I$ drops by roughly two orders of magnitude as $V_{dc}$ is decreased by one order of magnitude. $\nu = 1$ could not be measured for the complete voltage range, since current leakage across the facet ($V_{cc}\neq V_{cc'}$) became evident at higher voltages.

To analyze evidence for the conductance mechanism in the voltage dependence, in Fig. \ref{Vdep} we compare semi-logarithmic plots of the current $I$ and the conductance $G = I/V_{dc}$ vs. the inverse square root of the dc electric field ${\cal E}$. The electric field was obtained by assuming a uniform ${\cal E}$ along the junction and dividing the dc bias $V_{dc}$ by the junction length of 4 mm.
Nattermann, Malinin {\it et al.} \cite{Nattermann} predict one-dimensional variable-range hopping to have an exponential dependence $I({\cal E}) \sim \exp(-\sqrt{{\cal E}_0/{\cal E}})$ for electric fields $\cal E$ larger than the crossover field associated with the localization length. For $\nu =1$ ($\nu =2$), a fit with ${\cal E}_0 = 6.81$ V/cm (4.60 V/cm) describes the $I({\cal E})$ dependence below ${\cal E}^{-1/2} = 10\,{\rm \sqrt{V/cm}}$ (${\cal E}^{-1/2} =12.5\,{\rm \sqrt{V/cm}}$) quite well, as shown in Fig. \ref{Vdep} b). 
Fogler and Kelly \cite{Fogler_hopping} point out that Refs. \cite{Nattermann} do not correctly account for highly resistive segments in the conducting path, which are unavoidable in one dimension. They predict a linear response at low fields, followed by an intermediate regime, and an exponential dependence $G({\cal E}) \sim \exp(-\sqrt{{\cal E}_0/{\cal E}})$ for the conductance at high electric fields, with a typically s-shaped $I(V_{dc})$ curve. Fig. \ref{Vdep} c) shows a semi-logarithmic plot of the conductance $G$ vs. ${\cal E}^{-1/2}$, where only the $\nu = 2$ data at high fields above ${\cal E} = 0.25 \, {\rm V/cm}$ fits to an exponential behavior with ${\cal E}_0 = 1.01$ V/cm.
Our experimental data fits better to the earlier model \cite{Nattermann}, and we point out that the bent quantum Hall junction possibly represents a geometry where the highly resistive breaks on the 1D conducting path of Fogler and Kelly can be avoided: a highly resistive impurity could locally separate the chiral quantum Hall edges, and the current would circumvent the break instead of being strongly backscattered by it.   

It has been proposed \cite{Matt} that the insulating state could either be caused by a band gap in the dispersion, leading to activated conduction, or by localization leading to hopping conduction. While previous work \cite{Matt} was not able to distinguish between the two possible conduction mechanisms, the data presented here favors hopping conduction. The temperature and dc bias dependence are both consistent with 1D VRH conduction at low temperatures. The temperature dependence of the BQH junction conductance at $\nu = 1,2$ is consistent with a crossover from activated nearest-neighbor hopping conduction to 1D VRH conduction $G(T) \sim \exp(-(T_0/T)^{1/2})$ below $T = 320$ mK. The measured $I({\cal E})$ dependence is described best by 1D VRH conduction of the form $I \sim \exp(-{\cal (E}_0/{\cal E})^{1/2})$ \cite{Nattermann}. The larger value for ${\cal E}_0$ observed at $\nu = 1$ suggests a magnetic field dependent localization length, which decreases at higher fields.

We observe almost identical behavior for $\nu = 1$ and $\nu = 2$, although the bulk $\nu = 2$ state corresponds to a fully occupied, spin-degenerate lowest Landau level, while in the bulk $\nu = 1$ state the spin degeneracy is lifted. This spin-independent conductance is also observed in weakly insulating states at filling factors $3,4$ and $5,6$ \cite{Matt}. 
Experimentally, Kang {\it et al.} \cite{KANG} measured a zero-bias peak in the tunnel conductance across a short $15 \, \mu$m junction between lateral quantum Hall systems at $\nu = 1,2$ \cite{Kang_clarifier}, whereby the features at $\nu = 1,2$ also share surprisingly similar behavior.  To understand this similarity, it is first important to note that at such sharp-edge profiles as these, the formation of incompressible strips is suppressed \cite{Zozo}, leading to a very similar edge structure at different spin states of the same Landau level.
A further explanation for the $\nu = 1,2$ similarity has already been proposed by Kim and Fradkin \cite{kim}, who modeled a line junction of counter-propagating quantum Hall edges as a Luttinger liquid, where the effective Luttinger parameter $K$ is reduced by Coulomb interactions between forward and reverse movers. For $K < 1$ tunneling becomes a relevant perturbation and the conduction along the line junction is suppressed. For the fully spin-polarized ($\nu = 1$) and partially spin-polarized states ($\nu = 2$ with Zeeman coupling), it is expected that $K <1$ and both states should strongly backscatter and become insulating in the presence of disorder.

In conclusion, we performed magneto-transport studies of a bent quantum Hall junction with equal integer filling factors $\nu = 1,2$. The temperature and electric field dependence of the conductance suggest VRH conduction as the low-temperature transport mechanism. The qualitative temperature- and voltage-dependence is consistent with VRH conduction without interactions.  We observe a quantitatively similar temperature dependence at $\nu = 1$ and $\nu = 2$, analogous to previous measurements of tunnel-coupled quantum Hall edges by Kang {\it et al} \cite{KANG} which also showed qualitative similarities.  Following Kim and Fradkin \cite{kim}, it is suggested that interactions may nonetheless be responsible for enhancing the probability for backscattering along the junction, and this mechanism may partially explain the similarity between $\nu = 1$ and $2$.  It will be interesting to measure the conductance along shorter BQH junctions, better comparable to the $15 \mu$m used by Kang {\it et al.}

\begin{acknowledgments}
The authors thank Eun-Ah Kim, Dmitry Polyakov and Thierry Giamarchi for illuminating discussions and comments. The measurements were partly performed at the Max Planck Institut f\"ur Festk\"orperforschung, in collaboration with J. Smet, L. Hoeppel, and K. von Klitzing, and partly performed at the Grenoble High Magnetic Field Laboratory CNRS with D. Maude under EEC funding contract No. RITA-CT-2003-505474.
This work was supported by the Deutsche Forschungsgemeinschaft in the Schwerpunktprogramm Quanten Hall Systeme and by DFG GR 2618/1-1.
\end{acknowledgments}


\begin{thebibliography}{00}
\bibitem{GangOfFour} E. Abrahams, P. W. Anderson, D. C. Licciardello, and T. V. Ramakrishnan, Phys. Rev. Lett. {\bf 42}, 673 (1979).
\bibitem{SHK} B. I. Shklovskii and A. L. Efros, {\it Electronic Properties of Doped Semiconductors} (Springer, New York, 1984); B. I. Shklovskii, Fiz. Tekh. Poluprov. {\bf 6}, 2335 (1973) - Engl. transl.: Sov. Phys. Semicond. {\bf 6}, 1964 (1973). 
\bibitem{Fogler_hopping} M. M. Fogler and R. S. Kelley, Phys. Rev. Lett. {\bf 95}, 166604 (2005).
\bibitem{Nattermann} T. Nattermann, T. Giamarchi, and P. Le Doussal, Phys. Rev. Lett. {\bf 91}, 056603 (2003); S. V. Malinin, T. Nattermann, and B. Rosenow, Phys. Rev. B 70, 235120 (2004).
\bibitem{GOR} I. V. Gornyi, A. D. Mirlin, and D. G. Polyakov, Phys. Rev. Lett. {\bf 95}, 206603 (2005).
\bibitem{Basko} D. M. Basko, I. L. Aleiner, and B. L. Altshuler, Ann. Phys. (N. Y.) {\bf 321}, 1126 (2006). 
\bibitem{Halperin} B. I. Halperin, Phys. Rev. B {\bf 25}, 2185 (1982).
\bibitem{Wen} X. G. Wen, Phys. Rev. Lett. {\bf 64}, 2206 (1990).
\bibitem{KANG} W.Kang, H. L. Stormer, L. N. Pfeiffer, K. W. Baldwin and K. W. West, Nature {\bf 403}, 59 (2000); M. Habl, M. Reinwald, W. Wegscheider, M. Bichler, and G. Abstreiter, Phys. Rev. B {\bf 73}, 205305 (2006).
\bibitem{1} M. Grayson, D. Schuh, M. Huber, M. Bichler, and G. Abstreiter, Appl. Phys. Lett. {\bf 86}, 032101 (2005); M. Grayson, D. Schuh, M. Bichler, M. Huber, G. Abstreiter, L. Hoeppel, J. Smet, and K. v. Klitzing, Physica E {\bf 22}, 181 (2004); D. Schuh, M. Grayson, M. Bichler and G. Abstreiter, Physica E {\bf }{\bf 23}{\bf }, 293 (2004);
\bibitem{Matt} M. Grayson, L. Steinke, D. Schuh, M. Bichler, L. Hoeppel, J. Smet, K. v. Klitzing, D. K. Maude, and G. Abstreiter,  Phys. Rev. B 76, 201304(R) (2007).
\bibitem{footnote} The sample is the same as sample a) in Ref. \cite{Matt}.
\bibitem{RENN} S. R. Renn and D. P. Arovas, Phys. Rev. B {\bf 51}, 16832 (1995).
\bibitem{KANE_FISHER} C. L. Kane and M. P. A. Fisher, Phys. Rev. B {\bf 46}, 15 233 (1992).
\bibitem{KANE} C. L. Kane and M. P. A. Fisher, Phys. Rev. B {\bf 56}, 15231 (1997).
\bibitem{kim} E. A. Kim and E. Fradkin, Phys. Rev. B {\bf 67}, 045317 (2003).  
\bibitem{kollar} M. Kollar and S. Sachdev, Phys. Rev. B {\bf 65}, 121304(R) (2002).
\bibitem{mitra} A. Mitra and S. M. Girvin, Phys. Rev. B {\bf 64}, 041309(R) (2001).
\bibitem{ALE04} A. N. Aleshin, J. Y. Lee, S. W. Chu, S. W. Lee, B. Kim, S. J. Ahn, and Y. W. Park, Phys. Rev. B {\bf 69}, 214203 (2004); A. N. Aleshin, H. J. Lee, Y. W. Park, and K. Akagi, Phys. Rev. Lett. {\bf 93}, 196601 (2004).
\bibitem{Shante} V. K. S. Shante, C. M. Varma and A. N. Bloch, Phys. Rev. B {\bf 8}, 4885 (1973).
\bibitem{Kurki} J. Kurkij\"arvi, Phys. Rev. B {\bf 8}, 922 (1973).
\bibitem{Kang_clarifier} Note that if $G_{tunn}$ represents the tunnel conductance measurements of Kang {\it et al.}, then the conductance along the junction $G$ in the present paper can be related to the coupling conductance across the junction $G_{tunn}$ by $G = \sigma_{xy} - G_{tunn}$.
\bibitem{Zozo} S. Ihnatsenka and I. V. Zozoulenko, Phys. Rev. B {\bf 73}, 155314 (2006). 
\end{thebibliography}
\end{document}